\documentclass[conference]{IEEEtran}
\IEEEoverridecommandlockouts
\usepackage{pifont}
\usepackage{booktabs}
\usepackage{cite}
\usepackage{amsmath,amssymb,amsfonts}
\usepackage{algorithmic}
\usepackage{graphicx}
\usepackage{tabularx}
\usepackage{textcomp}
\usepackage{xcolor}
\usepackage[table]{xcolor}  
\usepackage{colortbl}       
\usepackage{hyperref}
\hypersetup{hidelinks}
\usepackage[ruled,vlined]{algorithm2e}
\def\BibTeX{{\rm B\kern-.05em{\sc i\kern-.025em b}\kern-.08em
    T\kern-.1667em\lower.7ex\hbox{E}\kern-.125emX}}
\begin{document}

\title{DFM: Difference Feature Modeling with Text-Guided Gated Contrastive Loss
for Remote Sensing Image Change Captioning}

\author{
\IEEEauthorblockN{
Yelin Wang\textsuperscript{1,*},
Zijia Song\textsuperscript{2,*},
Chuanguang Yang\textsuperscript{3,\dag},
Miaoyu Wang\textsuperscript{4},
Zhulin An\textsuperscript{3,\dag},
Libo Huang\textsuperscript{3},
and Yongjun Xu\textsuperscript{3}
}

\IEEEauthorblockA{
\textsuperscript{1}ShanghaiTech University \\
\textsuperscript{2}National University of Defense Technology \\
\textsuperscript{3}Institute of Computing Technology, Chinese Academy of Sciences \\
\textsuperscript{4}Goldman Sachs
}

\IEEEauthorblockA{
wangyl2023@shanghaitech.edu.cn, songzijia@nudt.edu.cn, yangchuanguang@ict.ac.cn,\\
wangrainm@gmail.com, anzhulin@ict.ac.cn, huanglibo@ict.ac.cn, xyj@ict.ac.cn
}

\IEEEauthorblockA{
\textsuperscript{*}Equal contribution \quad
\textsuperscript{\dag}Corresponding authors
}
}

\maketitle

\begin{abstract}
The primary goal of Remote Sensing Image Change Captioning (RSICC) is to automatically generate descriptions of changes between remote sensing images captured at different time points. Existing models still rely on a single autoregressive generation paradigm, which tends to prioritize learning easily generated vocabulary over capturing discriminative differences between images. To address this, we reframe the training paradigm and propose a novel Difference Feature Modeling (DFM) framework. Specifically, we introduce a Text-guided Gated Contrastive Loss (TGCL) to guide the vision encoder to extract critical features from a text-modal perspective. Additionally, we incorporate a pre-trained Change Detection model to transfer stable change detection knowledge. In order to further enhance the representation, we design a Joint Feature Modeling (JFM) module to achieve the fusion of multi-scale difference representations, thereby capturing comprehensive spatiotemporal variations between multi-temporal images. Extensive experiments on multiple datasets demonstrate the effectiveness of our approach.
\end{abstract}


\section{Introduction}
\label{sec:intro}
Remote Sensing Image Change Captioning (RSICC) is a critical task in the field of remote sensing . It involves analyzing multi-temporal remote sensing images of the same area and describing surface changes through natural language. Compared to Remote Sensing Image Change Detection (RSICD), which only requires segmenting the changed regions between multi-temporal images, RSICC leverages natural language to provide a more detailed characterization of changed object categories, positional relationships between objects, and dynamics of changing objects (e.g., additions or disappearances). RSICC significantly enhances the interpretability of change-related information and holds substantial application potential in various domains, including land planning, surface dynamics monitoring, and urban expansion research.

\begin{figure}[t]
  \centering
   \includegraphics[width=0.97\linewidth]{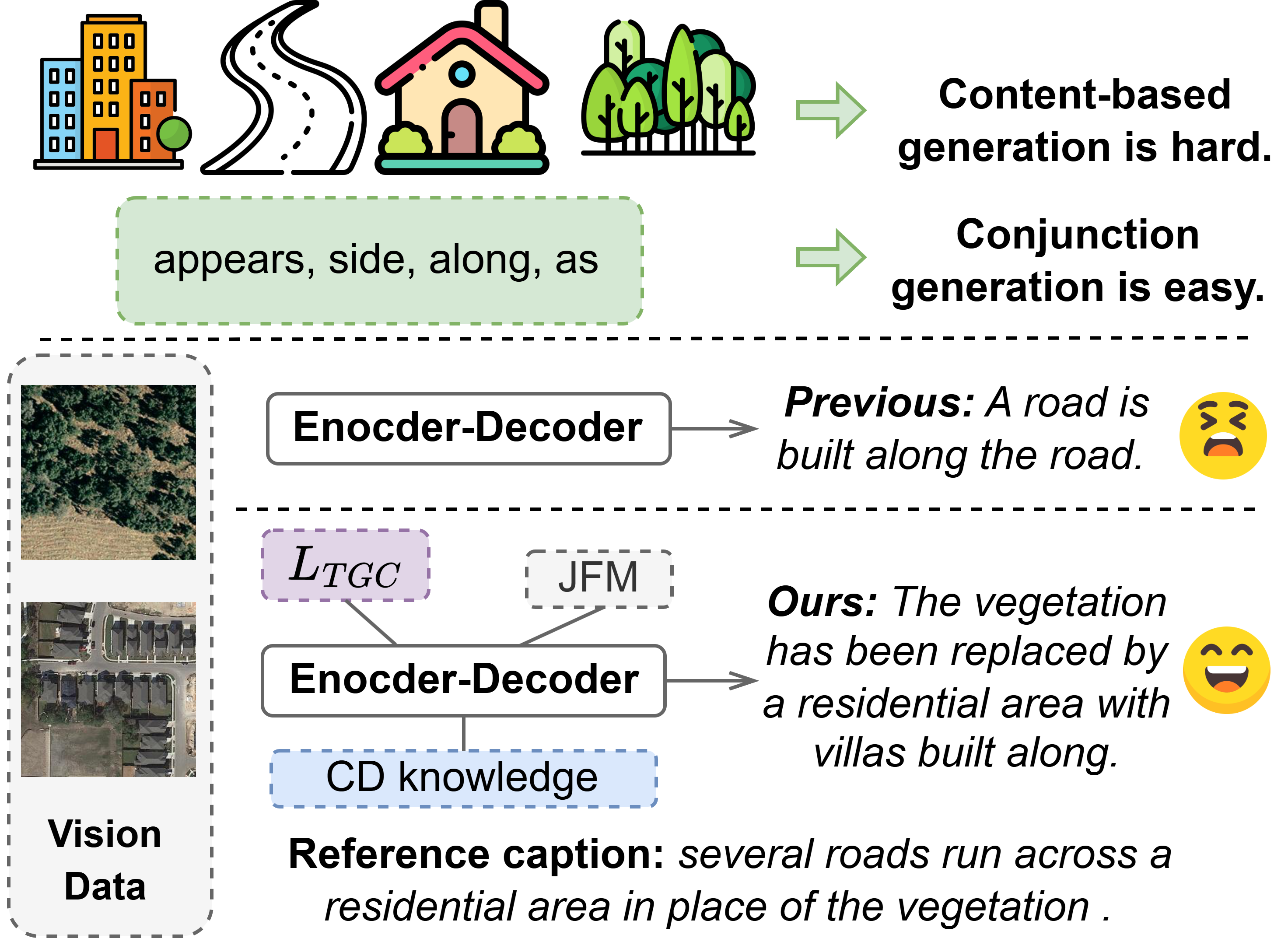}
   \caption{Challenges of previous models for RSICC and brief display of our method for generating semantically precise captions.}
   \label{intro}
   \vspace{-0.5cm}
\end{figure}

Compared to conventional image captioning, RSICC is a more challenging multi-modal task. It requires not only accurate interpretation of image content but also the ability to capture changes between multi-temporal images and generate corresponding natural language descriptions. Current mainstream methods \cite{RSCaMa,RSICC,SFA,chen2024high} typically adopt an encoder-decoder architecture, where visual features are encoded and decoded into textual descriptions. However, the critical challenge lies in guiding the model to effectively represent the visual changes between images while ensuring that these representations are sufficiently informative for the text decoder to generate semantically precise captions \cite{zhan2017change,liu2016deep,kerner2019toward,lin2019multispectral,gong2015change}.


To better characterize image changes, prior work explores change-aware captioning \cite{chouaf2021captioning,hoxha2022change}. PromptCC \cite{RSICC} decouples detection and description and adopts a multi-prompt strategy \cite{pierrat2025proximal,fassnacht2024remote,victor2024remote}. RSCaMa \cite{RSCaMa} applies State Space Models (SSM) for iterative spatial perception and temporal interaction. However, as shown in Figure~\ref{fig:cap-analysis}, these methods still rely on supervised autoregressive generation: captions are produced by a text decoder and optimized with a single loss across the vision encoder, change-focused neck, and decoder, which can induce model laziness. Moreover, the autoregressive loss uniformly supervises all tokens, ignoring position-dependent generation difficulty \cite{liu2023progressive,chen2024rsmamba}.

\begin{figure*}
  \centering
  \includegraphics[width=\textwidth]{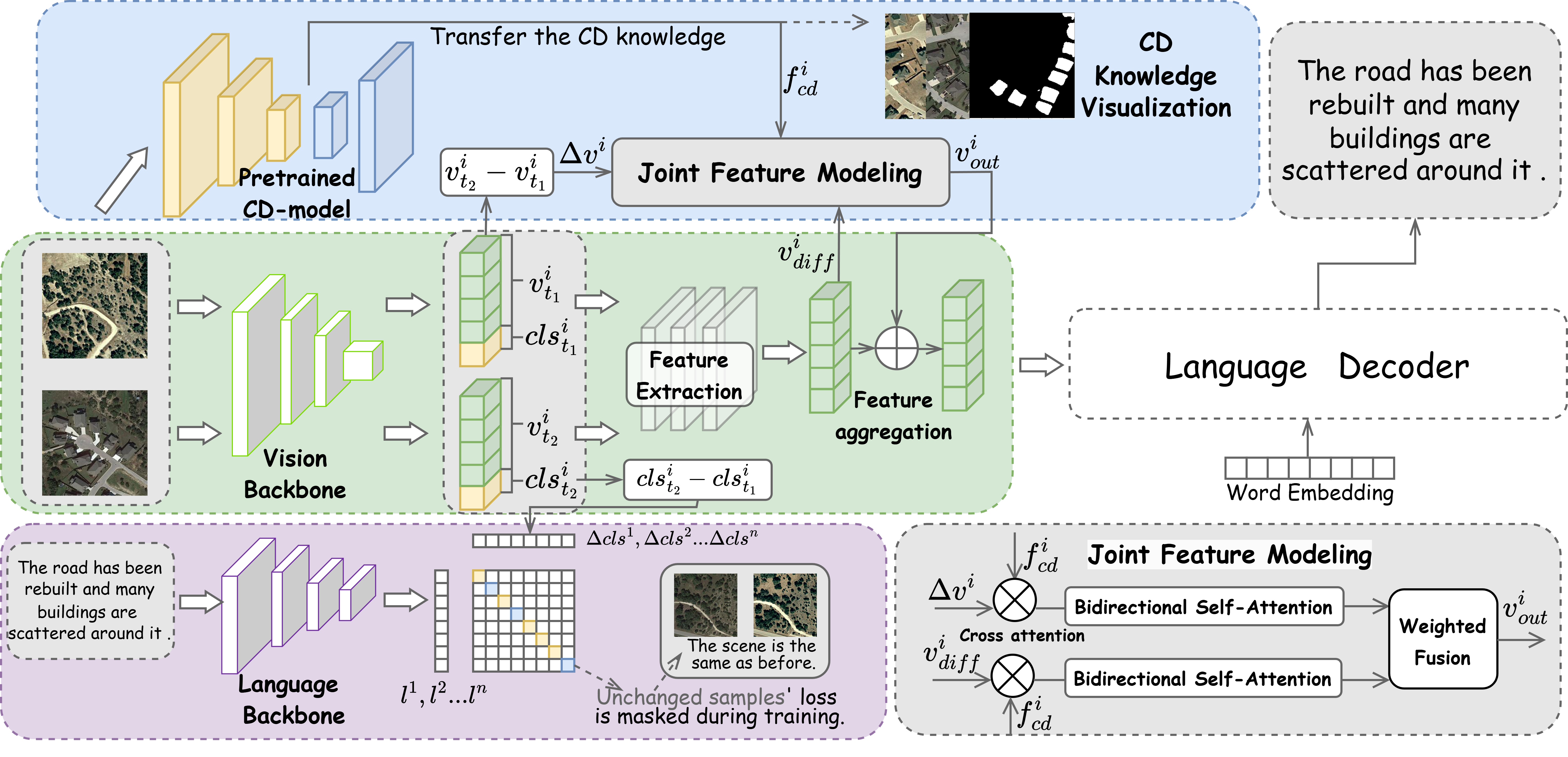}
  \caption{Flowchart of Difference Feature Modeling (DFM) framework for RSICC. A novel Text-Guided Gated Contrastive Loss is proposed, and the change detection knowledge is transferred through new designed Joint Feature Modeling module.}
  \label{fig:pipeline}
  \vspace{-0.5cm}
\end{figure*}

To address these challenges, we propose Difference Feature Modeling (DFM). Beyond the generation paradigm, DFM introduces cross-modal contrastive learning and integrates change detection knowledge to emphasize visually significant changes. Specifically, we add a text encoder and design a Text-guided Gated Contrastive Loss (TGCL) to align change-descriptive captions with difference features from multi-temporal images. A gating mechanism filters out “no-change” captions, preventing them from misleading optimization. In addition, we transfer spatial knowledge from pre-trained Change Detection models to strengthen change-related representations, and design a Joint Feature Modeling module (JFM) to fuse multi-scale features. By directly applying TGCL to the vision encoder, DFM enforces stronger constraints and drives the model toward visually grounded changes.
The main contributions of this work are summarized as follows:
\begin{itemize}
  \item \textbf{We identify model laziness in existing RSICC methods}, where autoregressive generation tends to over-rely on generic linguistic patterns instead of visual evidence. To mitigate this issue, we propose a \textbf{Text-guided Gated Contrastive Loss} and leverage pre-trained Change Detection models to enforce visually grounded change understanding.

  \item \textbf{We introduce a Joint Feature Modeling  module} that jointly captures pixel-level and structural difference features across multiple scales, enabling more accurate and context-aware change captioning.

  \item \textbf{Extensive experiments on multiple benchmarks} demonstrate the effectiveness of the proposed \textbf{Difference Feature Modeling} framework, achieving consistent improvements over state-of-the-art methods, particularly in terms of \textbf{CIDEr-D}.
\end{itemize}



\section{Related Works}
\subsection{Remote Sensing Change Detection}
Remote sensing change detection aims to identify changes between multi-temporal images and generate pixel-level change masks \cite{cheng2024change,wang2024advances,dong2024changeclip}. 
Early approaches mainly relied on pointwise classification and handcrafted features \cite{bovolo2008novel,seo2018fusion,huo2016learning,li2014sar}. 
With the advent of deep learning, CNN-based methods substantially improved change detection performance due to their strong representation and nonlinear modeling capabilities \cite{zhan2017change,liu2016deep,kerner2019toward}. 
More recently, attention mechanisms and transformer architectures have been introduced to capture spatial--temporal dependencies, achieving further performance gains \cite{vaswani2017attention,wang2021attention,li2022transunetcd,ding2022cdformer}. 
In parallel, emerging paradigms such as diffusion models and Mamba architectures have driven rapid progress in change detection research \cite{dhariwal2021diffusion,chen2024changemamba,liu2025vmamba}. 
Closely related to change detection, change captioning focuses on describing changes at the semantic level rather than pixel-wise localization \cite{RSICC}.

\subsection{Remote Sensing Image Change Captioning}

Remote Sensing Image Change Captioning is a relatively new research direction in remote sensing.
Chouaf \textit{et al}. \cite{chouaf2021captioning} first explored RSICC on a private dataset using a CNN--RNN encoder--decoder framework.
Subsequent works extended this paradigm by introducing different fusion strategies and architectures, including feature-level fusion and transformer-based models \cite{hoxha2022change,liu2022remote}.
Notably, Liu \textit{et al}. \cite{liu2022remote} released the large-scale LEVIR-CC dataset and established benchmark methods based on dual-branch transformer encoders.
Further improvements were achieved through multi-scale and difference-aware modeling to enhance the perception of changed objects \cite{liu2023progressive}.
More recently, Mamba-based architectures have been explored for dual-temporal spatiotemporal feature extraction \cite{chen2024rsmamba}.

Despite promising progress, existing RSICC methods still struggle with generating accurate, semantically rich, and detail-consistent change descriptions.
Motivated by these limitations, we propose a series of improvements to address the insufficient and inaccurate semantic modeling in prior work.

\section{Methodology}
\label{sec:method}

\subsection{Overview of Framework}
Previous methods predominantly adopt autoregressive generation, often leading to model laziness and repetitive captions rather than precise change descriptions.
To address this, we propose \textbf{Difference Feature Modeling}, which guides the model to focus on inter-image changes (Fig.~\ref{fig:pipeline}).
DFM introduces a \textbf{Text-guided Gated Contrastive Loss} to emphasize text-relevant change content and a \textbf{Joint Feature Modeling} module to refine change-descriptive representations using change detection priors.
The end-to-end training procedure is summarized in Alg.~\ref{alg:dfm}.

\subsection{Change Detection Branch}

We employ a pre-trained Change Detection (CD) model to extract spatially explicit change information from multi-temporal images and transfer this stable knowledge into vision representations to guide the language decoder.
Given an image pair, the CD model produces a binary change map highlighting changed regions.
Intermediate embeddings from the CD encoder are used as priors to guide the vision encoder, enhancing change-relevant features while suppressing background noise.


\begin{algorithm}[t]
\caption{Training of Difference Feature Modeling (DFM)}
\label{alg:dfm}
\KwIn{Mini-batch $\mathcal{B}=\{(X^i_{t1},X^i_{t2},c^i)\}_{i=1}^{B}$}
\KwIn{Frozen encoders: $\mathrm{CDModel\!-\!Enc}$, $\mathrm{TextEnc}$; Trainable: $\mathrm{ViT}$, $\mathrm{FeatureExtraction}$, JFM, $\mathrm{Decoder}$}
\KwIn{Loss weights $\alpha,\beta$}
\KwOut{Updated parameters $\theta$}

\For{$i=1$ \KwTo $B$}{
    \tcp{(1) Visual encoding}
    $v^i_{t1} \leftarrow \mathrm{ViT}(X^i_{t1}),\quad v^i_{t2} \leftarrow \mathrm{ViT}(X^i_{t2})$\;
    $v^i_{\mathrm{diff}} \leftarrow \mathrm{FeatureExtraction}(v^i_{t1}, v^i_{t2})$ 
    $\Delta v^i \leftarrow v^i_{t2}-v^i_{t1}$ 

    \tcp{(2) Transfer CD knowledge (frozen)}
    $f^i_{\mathrm{cd}} \leftarrow \mathrm{CDModel\!-\!Enc}(X^i_{t1}, X^i_{t2})$ 
    $a^i \leftarrow \mathrm{CrossAttn}(\Delta v^i, f^i_{\mathrm{cd}})$ 
    $b^i \leftarrow \mathrm{CrossAttn}(v^i_{\mathrm{diff}}, f^i_{\mathrm{cd}})$ 
    $v^i_{\mathrm{out}} \leftarrow \mathrm{WF}(\mathrm{BiAttn}(a^i),\ \mathrm{BiAttn}(b^i))$ 

    \tcp{(3) Text encoding (frozen) and gating for TGCL}
    $l^i \leftarrow \mathrm{Proj}(\mathrm{TextEnc}(c^i))$\;
    $\Delta cls^i \leftarrow cls(v^i_{t2}) - cls(v^i_{t1})$ \tcp*{CLS difference for TGCL}
    $\delta_i \leftarrow \mathbb{I}[\text{$i$-th pair has change}]$,\quad
    $N_{\mathrm{valid}} \leftarrow \sum_{k=1}^{B}\delta_k + \epsilon$ 
}

\tcp{(4) Build similarity matrix and TGCL}
\For{$i=1$ \KwTo $B$}{
\For{$j=1$ \KwTo $B$}{
$S_{i,j} \leftarrow \exp(\alpha)\cdot
\Big(\frac{\Delta cls^i}{\|\Delta cls^i\|_2}\cdot\frac{l^j}{\|l^j\|_2}\Big)$ 
}}
$L_{\text{text2img}} \leftarrow \frac{1}{N_{\mathrm{valid}}}\sum_{i=1}^{B}\delta_i\cdot \mathrm{CE}(S_i, i)$ 
$L_{\text{img2text}} \leftarrow \frac{1}{N_{\mathrm{valid}}}\sum_{i=1}^{B}\delta_i\cdot \mathrm{CE}(S_i^\top, i)$ 
$L_{\mathrm{TGC}} \leftarrow \frac{1}{2}(L_{\text{text2img}}+L_{\text{img2text}})$ 

\tcp{(5) Autoregressive caption generation loss}
$\hat{y}^i \leftarrow \mathrm{Decoder}(v^i_{\mathrm{out}}, w)$,\quad
$L_{\mathrm{gen}} \leftarrow \sum_{i=1}^{B}\mathrm{CE}(\hat{y}^i, y^i)$ 

\tcp{(6) Optimize}
$L \leftarrow \alpha L_{\mathrm{TGC}} + \beta L_{\mathrm{gen}}$\;
Update $\theta \leftarrow \theta - \eta \nabla_{\theta} L$ (freeze $\mathrm{CDModel\!-\!Enc}$ and $\mathrm{TextEnc}$)\;

\end{algorithm}



\subsection{Text-guided Gated Contrastive Loss}


Traditional encoder-decoder frameworks mainly rely on autoregressive loss, leading to inefficient optimization and a bias toward easy-to-generate frequent words rather than change-grounded captions.
From a contrastive learning perspective, image differences and their textual descriptions can be treated as paired data.
Aligning change-representative visual embeddings with corresponding caption embeddings encourages the vision encoder to focus on text-described changes.

\begin{table*}
  \centering
   \small  
  \normalsize
  \setlength{\tabcolsep}{4pt} 
    \caption{Performance comparison on the LEVIR\_CC dataset. Best results are in \textbf{bold}.}
     \vspace{-0.3cm}
  \begin{tabularx}{\textwidth}{@{}lXrccccccccc@{}} 
    \toprule
    Type &   Method & CIDEr-D & METEOR & ROUGE$_{L}$ & BLEU-1 & BLEU-2 & BLEU-3 & BLEU-4 & $S_m^*$ \\
    \midrule
    CNN based 
    & Capt-Rep-Diff \cite{Park2019RobustChangeCaptioning} & 110.57 & 34.47 & 65.64 & 72.90 & 61.98 & 53.62 & 47.41 & 64.52 \\
    & Capt-Att \cite{Park2019RobustChangeCaptioning} & 121.22 & 36.58 & 69.73 & 77.64 & 67.40 & 59.24 & 53.15 & 70.17 \\
    & Capt-Dual-Att \cite{Park2019RobustChangeCaptioning} & 124.42 & 36.56 & 70.69 & 79.51 & 70.57 & 63.23 & 57.46 & 72.28 \\
    & DUDA \cite{Park2019RobustChangeCaptioning} & 124.32 & 37.15 & 71.04 & 81.44 & 72.22 & 64.24 & 57.79 & 72.58 \\
    \midrule
    Transformer based 
    & MCCFormer-S \cite{Qiu2021MCCFormers}& 120.39 & 36.17 & 69.46 & 79.90 & 70.26 & 62.68 & 56.68 & 70.68 \\
    & MCCFormer-D \cite{Qiu2021MCCFormers}& 124.44 & 37.29 & 70.32 & 80.42 & 70.87 & 62.86 & 56.38 & 72.11 \\
    & RSICCFormer \cite{9934924}& 134.12 & 39.61 & 74.12 & 84.72 & 76.27 & 68.87 & 62.77 & 77.65 \\
    & PSNet \cite{liu2023progressive} & 132.62 & 38.80 & 73.60 & 83.86 & 75.13 & 67.89 & 62.11 & 76.78 \\
    & PromptCC \cite{RSICC}& 136.44 & 38.82 & 73.72 & 83.66 & 75.73 & 69.10 & 63.54 & 78.13 \\
    \midrule
    Mamba based 
    & RSCaMa \cite{RSCaMa} & \underline{136.56} & \underline{39.91} & \underline{75.24} & \underline{85.79} & \underline{77.99} & \underline{71.04} & \underline{65.24} & \underline{79.24} \\
    \midrule
    \rowcolor{yellow!20}
    Mixed
     & DFM(Ours) & \textbf{142.51} & \textbf{40.95} & \textbf{75.90} & \textbf{86.19} & \textbf{78.82} & \textbf{72.21} & \textbf{66.26} & \textbf{81.40} \\
    \bottomrule
  \end{tabularx}
\vspace{-0.3cm}
  \label{tab:levir_cc_results}
\end{table*}

\begin{table*}
  \centering
  \small
  \setlength{\tabcolsep}{4pt} 
    \caption{Performance comparison on the Dubai\_CC dataset. Best results are in \textbf{bold}.}
     \vspace{-0.3cm}
   \normalsize
  \begin{tabularx}{\textwidth}{@{}lXrccccccccc@{}}
    \toprule
    Type  & Method    & CIDEr-D & METEOR & ROUGE$_{L}$ & BLEU-1 & BLEU-2 & BLEU-3 & BLEU-4 & $S_m^*$ \\
    \midrule
    CNN based 
    & DUDA  \cite{Park2019RobustChangeCaptioning}   & 62.78  & 22.05  & 48.34 & 58.82  & 43.59  & 33.63  & 25.39 &  39.64\\
    \midrule
    Transformer based 
    & MCCFormers-S \cite{Qiu2021MCCFormers} & 53.81 & 18.64 & 43.29 & 52.97  & 37.02  & 27.62  & 22.57 &  34.57\\
    & MCCFormers-D \cite{Qiu2021MCCFormers}  & 66.51 & 25.09 & 51.27 & 64.65  & 50.45  & 39.36  & 29.48 & 43.08 \\
    & RSICCformer \cite{9934924} & 66.54 & \underline{25.41} & \underline{51.96} & \textbf{67.92}  & \textbf{53.61}  & \textbf{41.37}  & \underline{31.28} & \underline{43.79}\\
    \midrule
    Mamba based
    & RSCaMa \cite{RSCaMa} & \underline{67.46} & 23.51 & 49.14 & 63.39 & 47.93  & 36.65  & 27.47 & 41.24 \\
    \rowcolor{yellow!20}
    \midrule
    Mixed
    & DFM(Ours)  & \textbf{84.65} & \textbf{25.75} & \textbf{54.36} & \underline{67.86} & \underline{51.01}  & \underline{40.61}  & \textbf{32.08} & \textbf{49.21}   \\
    \bottomrule
  \end{tabularx}
\vspace{-0.5cm}
  \label{tab:dubai_cc_results}
\end{table*}

The RSICC dataset contains many no-change descriptions (e.g., “The scene is the same as before”), which, if aligned with near-zero visual difference embeddings, would cause optimization conflicts. To address this, we introduce a gating mechanism to exclude no-change representations from contrastive learning. We adopt the [CLS] token from the vision encoder to capture semantic-level changes, and use the pre-trained All-MPNet-base-v2 \cite{siino2024all} as the text encoder for sentence-level semantics. A projection layer maps text embeddings into the visual latent space, with the text encoder frozen during training. Based on this design, we propose the Text-guided Gated Contrastive Loss (TGCL) and provide formal format as follows:
\begin{equation}
\mathcal{L}_{\text{TGC}} = \frac{1}{2} \left( \mathcal{L}_{\text{text2img}} + \mathcal{L}_{\text{img2text}} \right) 
\end{equation}
where the TGCL $\mathcal{L}_{\text{TGC}}$ consists of a bidirectional  symmetrical loss composed of text-to-image ${L}_{\text{text2img}}$ and image-to-text ${L}_{\text{img2text}}$, and the formats are defined as follows:
\begin{equation}
\mathcal{L}_{\text{text2img}} = \frac{1}{N_{\text{valid}}} \sum_{i=1}^B \delta_i \cdot \mathrm{CE}(\mathbf{S}_i, i) 
\end{equation}
\begin{equation}
\mathcal{L}_{\text{img2text}} = \frac{1}{N_{\text{valid}}} \sum_{i=1}^B \delta_i \cdot \mathrm{CE}(\mathbf{S}_i^\top, i)
\end{equation}
where $B$ is the batch size, $\mathrm{CE}(\cdot,\cdot)$ the cross-entropy loss, $\delta_i$ the gating indicator, and $N_{\text{valid}}$ the number of valid samples (see Eq.~\ref{number}). The similarity matrix \( S \in \mathbb{R}^{B \times B} \) is computed as:
\begin{equation}
\mathbf{S}_{i,j} = \exp(\alpha) \cdot \left( \frac{\Delta cls^i}{\left\|\Delta cls^i\right\|_2} \cdot \frac{l^j}{\left\|l^j\right\|_2} \right) 
\end{equation}

where $\alpha$ is a learnable logit scaling factor, $\Delta cls^i$ denotes the difference representation of the $i$-th image pair, and $l^j$ represents the $j$-th caption. The binary gating term $\delta_i \in \{0,1\}$ indicates whether a change occurs in the $i$-th image pair, and the number of valid samples $N_{\text{valid}}$ are defined as:
\begin{equation}
N_{\text{valid}} = \sum_{i=1}^B \delta_i + \epsilon 
\label{number}
\end{equation}
where \( \epsilon \) is a small constant which is usually set to \( 1 \times 10^{-8} \) to ensure numerical stability and avoid division by zero. 

Combined with the autoregressive generation loss $\mathcal{L}_\mathrm{gen}$, our overall optimization objective is formulated as follows:
\begin{equation}
    \mathop{\min}_{\theta} ~\alpha \mathcal{L}_\mathrm{TGC}+\beta \mathcal{L}_\mathrm{gen}
\end{equation}
where $\theta$ denotes all learnable parameters of DFM, and $\alpha$ and $\beta$ are hyperparameters balancing different objectives.
$\mathcal{L}_\mathrm{TGC}$ explicitly constrains the vision encoder to focus on caption-relevant changes, while $\mathcal{L}_\mathrm{gen}$ optimizes the model holistically.

\subsection{Joint Feature Modeling}

To better exploit differences between multi-temporal images, we design modeling techniques for image differences (Figure~\ref{fig:pipeline}). In the visual branch, two remote sensing images from distinct time points pass through a Vision Transformer, producing representations ${v_{t_1}^i, v_{t_2}^i}$, where $i$ denotes the image pair and $t_1, t_2$ the time points. These representations are processed by feature extraction modules to obtain enhanced spatiotemporal features $v_{\mathrm{diff}}^i$:
\begin{equation}
v_{\mathrm{diff}}^i = \mathrm{FeatureExtraction}(v_{t_1}^i,\, v_{t_2}^i)
\end{equation}
where $v_{\mathrm{diff}}^i$ captures the difference between multi-temporal images. To further refine and enhance the modeling of image differences, we design a dedicated \textbf{Joint Feature Modeling} module. This module first computes direct difference features $\Delta v^i$ from the raw ViT embeddings:
\begin{equation}
\Delta v^i = v_{t_2}^i - v_{t_1}^i
\end{equation}
here, to leverage external prior knowledge from pretrained Change Detection models, we extract the output embeddings $f_{\mathrm{cd}}^i$ of the CD model’s encoder:
\begin{equation}
f_{\mathrm{cd}}^i = \mathrm{CDModel-Enc}(X_{t_1}^i, X_{t_2}^i)
\end{equation}
here, $X_{t_1}^i$ and $X_{t_2}^i$ are input images at time points $t_1$ and $t_2$, and $\mathrm{CDModel\text{-}Enc}$ denotes the encoder of the CD model. We inject the CD prior $f_{\mathrm{cd}}^i$ into both the direct-difference features $\Delta v^i$ and the spatiotemporal features $v_{\mathrm{diff}}^i$ via Cross-Attention. The resulting representations are then refined by Bidirectional Self-Attention, and finally fused through $\mathrm{WF}$ as:
\begin{equation}
\label{eq:wf_all}
\begin{aligned}
v_{\mathrm{out}}^i 
= \mathrm{WF}\Bigl(&
   \mathrm{BiAttn}\bigl(\mathrm{CrossAttn}(\Delta v^i,\; f_{\mathrm{cd}}^i)\bigr), \\
   & \mathrm{BiAttn}\bigl(\mathrm{CrossAttn}(v_{\mathrm{diff}}^i,\; f_{\mathrm{cd}}^i)\bigr)
 \Bigr)
\end{aligned}
\end{equation}

here, $\mathrm{WF}$ denotes Weighted Fusion, which balances the contributions of different streams via learnable weights.
The fused representation $v_{\mathrm{out}}^i$ is then fed into a language decoder together with word embeddings $w$ to generate the change caption $y^i$.

\begin{table}[t]
\caption{An ablation study of $\mathcal{L}_{\text{TGC}}$ on LEVIR\_CC. The \emph{Mask} column shows whether losses of unchanged samples are masked. $\Delta cls$ and $\Delta v$ indicate which vision backbone features are aligned under $\mathcal{L}_{\text{TGC}}$. The \emph{Encoder} column specifies the text encoder for feature extraction.
}
\vspace{-0.3cm}
\centering
\small
\setlength{\tabcolsep}{2pt} 
\resizebox{\columnwidth}{!}{%
\begin{tabular}{cccc|ccccc}
\toprule
$\mathcal{L}_{\text{TGC}}$ & Mask & $\Delta cls$/$\Delta v$ & Encoder & CIDEr-D & ROUGE$_{L}$ & METEOR & BLEU-4 & $S_m^*$ \\
\midrule
\ding{55} & -        & -          & -                 & 137.44 & 75.24 & 39.98 & 65.58 & 79.56 \\
\ding{51} & \ding{55} & $\Delta v$ & \textit{CLIP}     & 135.69 & 74.07 & 39.45 & 63.74 & 78.24 \\
\ding{51} & \ding{51} & $\Delta v$ & \textit{CLIP}     & 136.49 & 73.90 & 39.87 & 64.53 & 78.70 \\
\ding{51} & \ding{51} & $\Delta cls$ & \textit{CLIP}   & 140.51 & 75.33 & 40.38 & 65.06 & 80.32 \\
\ding{51} & \ding{51} & $\Delta cls$ & \textit{All-Mpnet} & 142.51 & 75.90 & 40.95 & 66.26 & 81.41 \\
\bottomrule
\end{tabular}%
}
\vspace{-0.3cm}
\label{tab:ablation1}
\end{table}

\begin{table}[t]
\caption{An ablation study of JFM on Dubai\_CC, comparing multi-scale feature fusion methods. \emph{\textbf{Gate}} uses gated fusion, \emph{\textbf{Concat}} concatenates features, \emph{\textbf{Bilinear}} applies a bilinear layer, \emph{\textbf{Weighted}} learns feature weights, and \emph{\textbf{Ours}} denotes our JFM fusion.}
\vspace{-0.3cm}
\centering
\resizebox{\columnwidth}{!}{%
\begin{tabular}{c|ccccc}
\toprule
Mechanism & CIDEr-D & ROUGE$_{L}$ & METEOR & BLEU-4 & $S_m^*$ \\
\midrule
Gate      & 68.752  & 52.175 & 23.574 & 28.541 & 43.26 \\
Concat    & 80.856  & 51.561 & 24.132 & 28.571 & 46.28 \\
Bilinear   & 76.679  & 52.420 & 24.537 & 26.262 & 44.97 \\
Weighted  & 78.136  & 53.312 & 24.920 & 28.430 & 46.20 \\
Ours      & 84.650  & 54.360 & 25.750 & 32.080 & 49.21 \\ 
\bottomrule
\end{tabular}
}
\label{tab:ablation2}
\vspace{-0.5cm}
\end{table}

\section{Experiments}

\begin{figure}[t]
  \centering
  \vspace{-0.5cm}
  \includegraphics[width=0.5\textwidth]{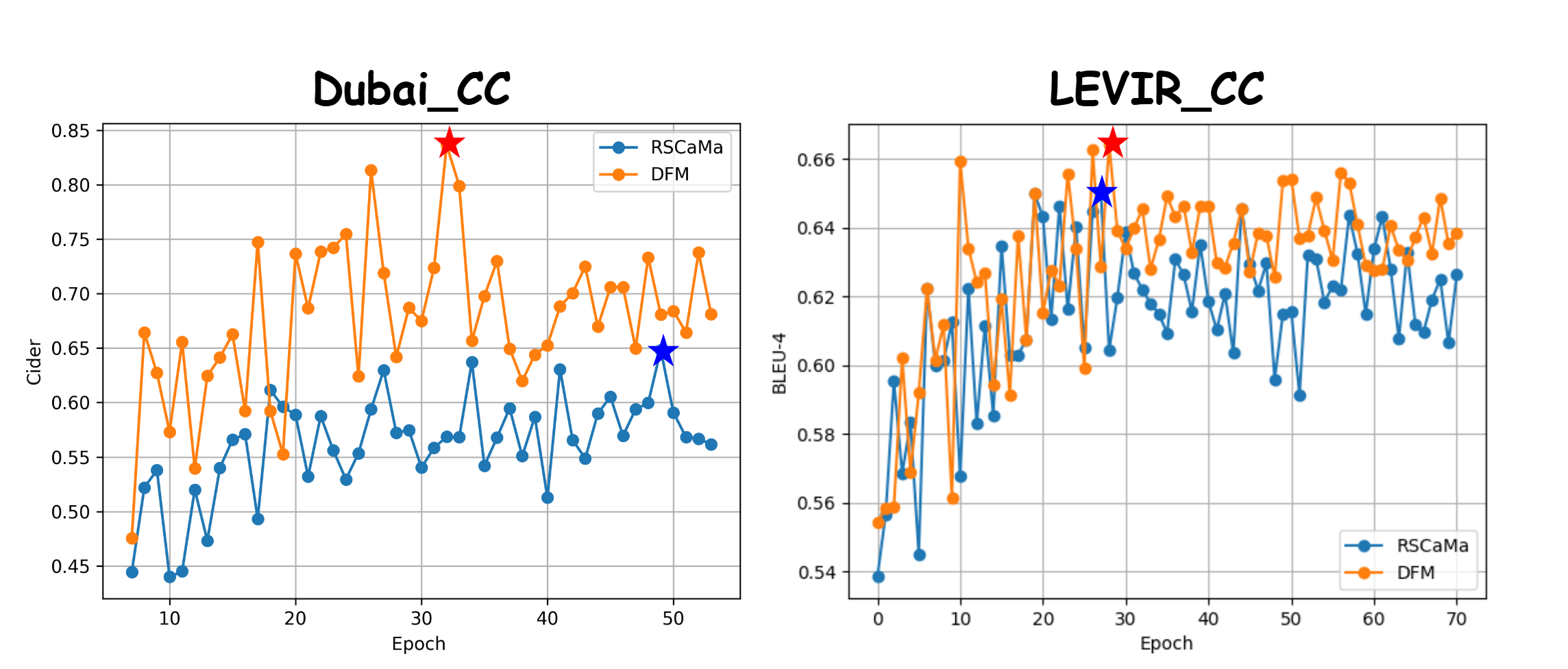}
  \vspace{-0.6cm}
  \caption{Line charts of BLEU-4 and Cider-D metrics over training epochs on different datasets. The red star represents the highest value of DFM, while the blue star represents the highest value of RSCaMa.}
  \label{fig:line}
  \vspace{-0.4cm}
\end{figure}

\begin{figure}[t]
  \centering
  \includegraphics[width=0.45\textwidth]{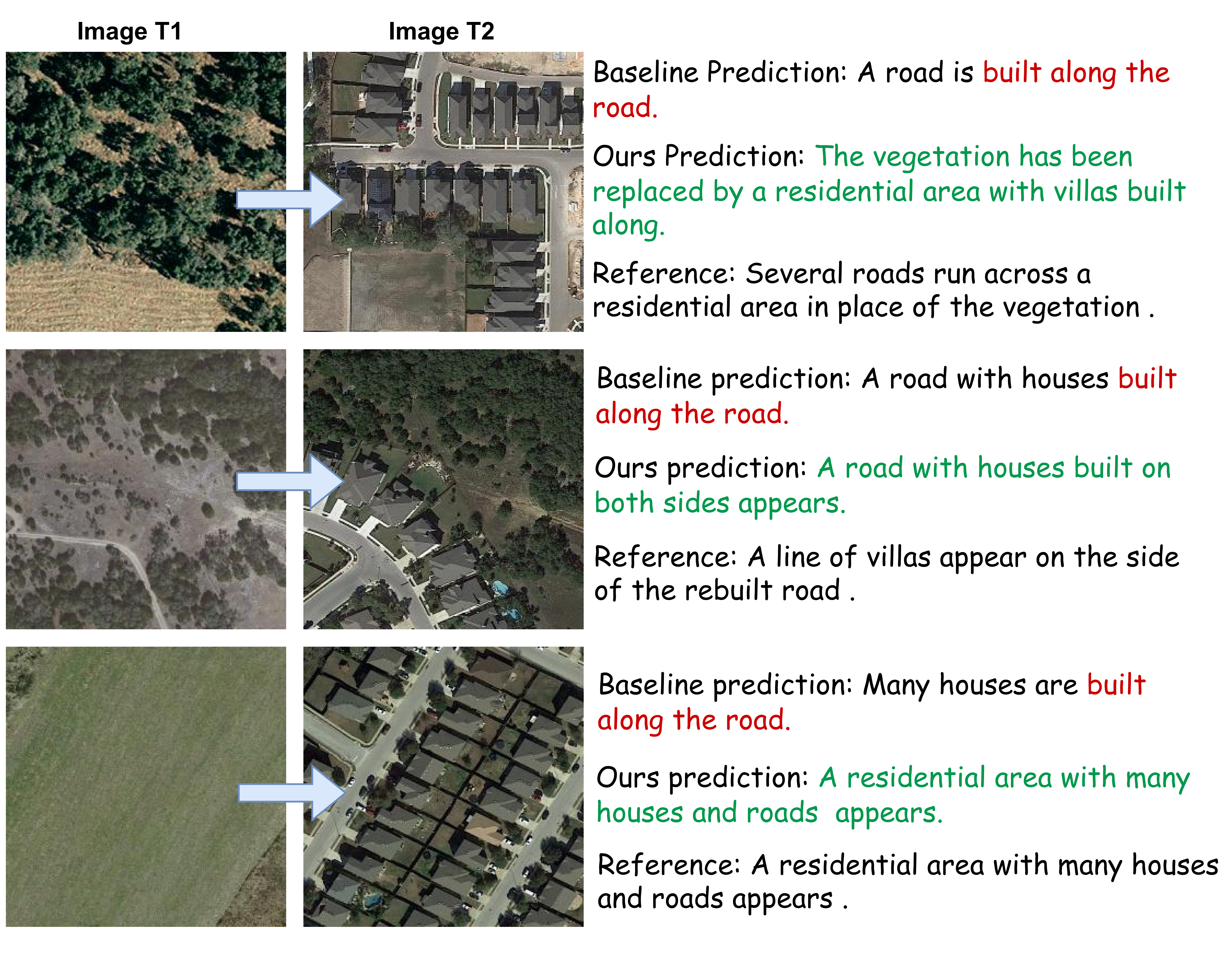}
  \caption{Qualitative comparison of baseline predictions, our method, and reference captions on LEVIR\_CC.}
  \label{fig:cap-analysis}
  \vspace{-0.5cm}
\end{figure}
\subsection{Experimental Setting}
\textbf{Datasets and Evaluation Metrics:} Our datasets include LEVIR-CC \cite{liu2022remote} and Dubai-CC \cite{hoxha2022change}. During training, we use Adam with a learning rate of $1\mathrm{e}{-4}$, batch size $64$, and embedding dimension $512$.  

%
\noindent\textbf{Evaluation Metrics:} We evaluate captioning accuracy using BLEU-1/2/3/4, ROUGE$_L$, METEOR, and CIDEr-D \cite{liu2025vmamba}, and also report the averaged score $S^*_m$ following \cite{RSCaMa}:  

\begin{equation}\small
S^*_m = \frac{1}{4} \left(\text{BLEU-4} + \text{ROUGE}_L + \text{METEOR} + \text{CIDEr-D}\right)
\end{equation}




\subsection{Comparison With the State-of-the-Art}
We compared our DFM with various state-of-the-art models based on different basic design components. Table~\ref{tab:levir_cc_results} presents the performance comparison on the LEVIR\_CC dataset. 
 Table~\ref{tab:dubai_cc_results} presents the comparison results on the Dubai\_CC dataset against existing excellent methods.

\subsection{In-depth Analysis of DFM}

\noindent\textbf{Comparison of training curves.}
Fig.~\ref{fig:line} compares the BLEU-4 and CIDEr-D training curves on Dubai\_CC and LEVIR\_CC, with RSCaMa as the baseline. DFM consistently outperforms RSCaMa throughout training, and the advantage is more pronounced on the smaller Dubai\_CC dataset, highlighting its stronger few-shot learning ability.




\noindent\textbf{Ablation Analysis of $\mathcal{L}_{\text{TGC}}$.}
We conduct an ablation of $\mathcal{L}_{\text{TGC}}$ on LEVIR\_CC (Table~\ref{tab:ablation1}). Row~1 is the baseline without $\mathcal{L}_{\text{TGC}}$. Applying it without masking (row~2) reduces to a standard contrastive loss and harms performance. Masking with $\Delta v$ (row~3) yields limited gains, while switching to $\Delta cls$ (row~4) brings clear improvements. Using All-Mpnet-base-V2 as encoder (row~5) achieves the best results due to stronger global semantic modeling.

\noindent\textbf{Ablation Analysis of Joint Feature Modeling.}
Multi-scale feature fusion is vital in JFM. As shown in Table~\ref{tab:ablation2}, gating and concatenation yield limited gains, while bilinear fusion improves some metrics but lowers BLEU-4. Weighted fusion stabilizes performance yet remains suboptimal. Our bidirectional self-attention with weighted fusion achieves the best results, confirming the effectiveness of our design.

\noindent\textbf{Qualitative Analysis of Experimental Results.}
Qualitative comparison shows that baseline methods rely on a supervised generation paradigm, where captions are produced by the text decoder and optimized with an autoregressive loss. This single loss may cause model laziness, as the forward pass involves the vision encoder, neck module, and text decoder but lacks diverse constraints. Moreover, the autoregressive loss applies uniform supervision to all tokens, ignoring differences in generation difficulty. As shown in Fig.~\ref{fig:cap-analysis}, simple words like “along” can be derived from context, while phrases such as “residential area” require visual grounding. As a result, the model tends to learn frequent patterns (e.g., “built along the road”) rather than capturing critical image changes.
\section{Conclusion}
In this paper, we propose a framework optimized by Text-guided Gated Contrastive Loss and enhanced with pretrained change detection features via the Joint Feature Modeling module. Experiments across multiple datasets show that our DFM generates more precise captions, achieving notable CIDEr-D improvements of 5.95\% on LEVIR\_CC and 17.19\% on Dubai\_CC.

\section*{Acknowledgment}
This work was supported by the National Natural Science Foundation of China under Grant Nos.~62406312 and 62476264.





\bibliographystyle{IEEEbib}
\bibliography{icme2026references}

\end{document}